\documentstyle[manuscript,aps,epsf]{revtex}

\begin{document}
\def\br{{\bf r}}
\def\fy{{\vec\phi}}
\def\valpha{{\vec\alpha}}
\def\vbeta{{\vec\beta}}
\draft
%\bibliographystyle{unsrt}
%\hfil{\today}
\title{ Probability distributions for multicomponent systems with multiplicative noise.}
\author{J.M. Deutsch}
\address{University of California\\
Santa Cruz CA 95064\\ }
\maketitle
%\vskip 1.5cm
\begin{abstract}
Linear systems with many degrees of freedom 
containing multiplicative and additive noise are considered.
The steady state probability distribution for equations of this kind
is examined. With multiplicative white noise it is shown that 
under some symmetry conditions, the probability distribution of a single
component has power law tails, with the exponent independent of the
strength of additive noise, but dependent on the strength of the multiplicative
noise. The classification of these systems into two regimes appears
to be possible in the same manner as with just one degree of freedom.
A physical system, that of a turbulent fluid undergoing
a chemical reaction is predicted to show a transition from exponential
to power law tails, as the reaction rate is increased.
A variety of systems are studied numerically. A replication algorithm
is used to obtain the Lyapunov exponents for high moments, which
would be inaccessible by more conventional approaches.
\end{abstract}
\newpage
\narrowtext
\section{INTRODUCTION}
\par
This paper analyzes coupled linear equations containing additive and
multiplicative noise. They can be written in the form
\begin{equation}
{\dot\fy} =  {\bf M\cdot\fy} + {\bf A \cdot\fy} + {\vec\eta}
\label{eq:vector}
\end{equation}
${\fy} ~\equiv ~ (\phi _1,\phi _2, \dots,\phi _N)$ represents the
variables of interest
${\bf M }(t)$ is a matrix that varies randomly with time.
$\bf A$ is a time independent matrix and $\vec\eta$ is a random
time dependent vector.
To motivate the study of eqn. (\ref{eq:vector})
we will illustrate a number of physical problems that are in this
category.

The motion of dye, or a temperature field, in a randomly stirred fluid
leads to an equation
for the dye density where the random velocity $\bf v$ is coupled to
dye density $\phi$ in a multiplicative way $\bf v \cdot \nabla \phi$.
obeying the equation
\begin{equation}
\partial_t \phi + {\bf v} \cdot \nabla \phi = d \nabla^2 \phi
\label{eq:passive}
\end{equation}
In two and higher dimensions this is normally studied
with the incompressibility condition $\nabla\cdot v ~=~ 0$.
$d$ is a diffusion constant, and $v~=~v({\bf r},t)$ is a random
function of position and time. This problem has received much
recent attention
~\cite{sinai,kraichnan,pope,eswaran,libchaber,sreenivasan,ching,yakhot,chen,orzag,pumir}.

The Schroedinger equation with a random time dependent potential
is another example~\cite{ovchinikov,girvin}, although it will not
be investigated here, and this is almost identical 
to the equation for wave propagation in random media~\cite{fock,martin}.
Population growth models which are relevant to chemical
reactions~\cite{nicolis},
population biology~\cite{zhang} and low temperature quantum
systems~\cite{koonin,mikhailov} are often of this form
\begin{equation}
\partial_t \phi ~=~ \alpha \phi + f \phi + d \nabla^2 \phi
\label{eq:population}
\end{equation}
where $f$ is a random function of position and time, $\alpha$  and
$d$ are parameters.  
Dye laser theory also leads to an equation with both types of noise~\cite{zhu}.
Polymers in turbulent flow also make use of such equations~\cite{armstrong,deutsch}.

The inclusion of additive noise most frequently corresponds to adding 
thermal fluctuations to the problem. In some cases it is a source
term and replaces the effect of a boundary. For example in turbulent
Raleigh-Bernard convection, where a liquid is heated from below, 
hot ``plumes"\cite{kadanoff}
form and rise up through a liquid. In this case $\phi$ in 
eqn. (\ref{eq:passive}) represents the temperature field, and a source term
in this equation should be added representing heat flow through the
boundaries.
\par
A one component problem similar to that studied here has been
analyzed by Drummond~\cite{drummond}. However his analysis 
is not completely applicable to the systems mentioned above as in
these cases the
additive and multiplicative noise are uncorrelated with each other.
However he also finds two regimes, one with all moments defined
and the other with a divergence at a finite moment.

The classification into different two regimes is very similar
to the one component case~\cite{one} where it was shown that
there there are two different types of behavior found for the
probability distribution function (PDF). There it was shown that
the PDF has power law tails in one regime and stretched exponential
in another.  In section \ref{sec:many} the n-component vector version
of these equations is considered
for the case where both kinds of noise are white and Gaussian. 
Here under
certain symmetry conditions it is shown that one also finds
a power law tail for the PDF of any
of the vector's components. 

An important example of non-power law tails involving many components
is eqn. (\ref{eq:passive}). $L(q)$ in this case should lead to
a stretched exponential or an exponential tail for the PDF 
of dye density. Such distributions have been
observed experimentally and have already been the subject
of much theoretical work. The argument presented here for these tails
is much general and just relies on mass conservation, that fact that
dye is conserved.

Is there a physical system similar to eqn. (\ref{eq:passive}) 
where the dye particles are not conserved? An example of such
a system would be a reactive fluid, where heat is generated
by particles reacting with each other. Such a system
would have Lyapunov exponents that cross over to the first
regime.  This leads to the interesting prediction that such
systems can exhibit power tails. This will be discussed in 
section \ref{sec:reaction}.
Numerical confirmation of this prediction is given in
section ~\ref{sec:numerics}.

\section{distribution in the absence of additive noise}
\label{sec:absence}
Here we will consider eqn. (\ref{eq:vector}) when the additive noise
term $\vec\eta $ is zero. In this case there is not a well defined
steady state probability distribution. However this case can also be understood
by considering the time dependence of the distribution for long times.
We shall see that there is a scaling form of this distribution for long times
which is identical in form to the one-component case.
\subsection{Scaling of the probability distribution}
Without the additive noise, the formal solution to eqn. (\ref{eq:vector}) is
\begin{equation}
\fy(t) ~=~ T e^{\int_0^t \bf M (\tau ) + \bf A d\tau } \fy (0)
\label{eq:formal}
\end{equation}
Here $T$ denotes the time ordered product.
If this integration is discretized into time intervals $\Delta t$,
then this is equivalent to the problem of the multiplication of 
$t/\Delta t$ random matrices of the form $\exp (\Delta t ({\bf M_i } +{\bf A}))$
The matrices $M_i ~=~ M(i \Delta t)$ are random and independent because we
are considering $M(t)$ to have a white noise spectrum.
\par
The multiplication of random matrices has been well studied. For long times
it has been proved that $|\fy (t)|$ has an exponential dependence on time for
long times~\cite{oseledec}. Fluctuations in this system are large in the
sense that higher moments do not scale in a trivial way with lower moments.
In one considers a single component of $\phi$, $\phi_1$ then if
$\phi_1 \sim \exp(\gamma_1 t)$ , then $\phi_1^2 \sim \exp(\gamma_2 t)$,
where $\gamma_1$ and $\gamma_2$ are 
not trivially related. This can be seen to be true even in the scalar case
where the matrices are dimensions $1 \times 1 $ and hence commute. 
In summary, one can  define a generalized
Lyapunov exponent $L(q)$ by~\cite{benzi,paladin}
\begin{equation}
\langle \phi_i^q\rangle \propto e^{L(q)t}
\label{eq:lyapunov}
\end{equation}
where the brackets denote an ensemble average over the noise $\bf M$.

Now we would
like to work out the form of the PDF of $\phi$,
one of the components of $\vec\phi$.
This situation is similar to the problem of multifractals and its solution works
out the same way. It can be checked that the distribution giving such scaling
is
\begin{equation}
\ln P(\ln \phi) \propto t f((\ln \phi) /t) + {\cal O}({\ln t \over t} )
\label{eq:plnphi}
\end{equation}
\par
Another way of deriving this scaling form is 
by a thermodynamic analogy~\cite{benzi,paladin}.
The discretized version of eqn. (\ref{eq:formal}) can be thought of as
a coupled  one dimensional spin system, where each spin has $N$ states
but with random interactions $\bf M_i$. Eqn. (\ref{eq:formal}) is then the partition
function of the system for a particular realization of the $\bf M_i$.
\par
From eqn. (\ref{eq:plnphi}) it is clear that $P (\phi)$ does not tend towards
a time independent, that is steady state, distribution. $\langle (\ln\phi )^2\rangle
\propto t$ for large $t$, so the the distribution continues to increase for
small and large $|\phi|$.

\section{Properties For Many Components}
\label{sec:many}
\subsection{Equation for moments}
\label{sec:moments}
\par
We will start by considering the coupled differential
eqns. (\ref{eq:vector})
where $\bf M$ is a random gaussian matrix with
$\langle M_{ij}
(t) M_{kl} (t')\rangle ~=~ 2\Gamma_{ijkl} \delta (t-t')$, 
and where $\bf A$ is taken
to be symmetric, real,  and time independent. As will be seen below,
the elements of $\bf A$ must be sufficiently negative so that
there is a well defined stationary solution, that is the solution
does not diverge for long times. $\vec\eta$ is additive random noise that is
taken to be white and uncorrelated $\langle\eta_i (t) \eta_j (t')\rangle
 ~=~ D\delta (t-t') \delta_{i,j}$. 
Later in this section we will see that when
${\bf M }(t)$ is either real symmetric or
anti-symmetric, the steady state
probability distribution for $\phi_i$ has a power law tail.
\par
Using standard methods~\cite{tatarskii} it is possible to derive the steady state
probability distribution for ${\fy}$.
\begin{equation}
[{\partial\over{\partial t}}+
\sum_i^N {\partial\over\partial \phi _i}(\sum_j A_{ij} \phi _j
+\sum_{jl} \Gamma_{ijjl} \phi _l) -\sum_{ik} {\partial^2 \over\partial \phi _i
\partial \phi _k}\sum_{jl} \Gamma_{ijkl} \phi _j \phi _l -\sum_i
D{\partial^2\over\partial \phi _i^2}] P(\fy ) ~=~0
\end{equation}
Next consider nth moments of this equation. Multiply through this equation
by $ \phi _{\alpha_1} \phi _{\alpha_2} \dots \phi _{\alpha_n} $ where the $\alpha_i$'s can
take on any integral value between $1$ and $N$. Integrating this over all
the $\phi _i$'s gives an equation relating moments of order $n$ to moments of
order $n-2$. 
It is convenient to write such an equation in matrix form.
Define $\valpha\equiv ( \alpha_1 ,\alpha_2 ,\dots ,\alpha_n )$ and
$\vbeta\equiv (\beta_1 , \beta_2 , \dots , \beta_n )$ where
the $\alpha_i$'s and $\beta_i$'s can
take on any integral value between $1$ and $N$.
Define
$\fy_{\valpha} \equiv \langle \phi _{\alpha_1} \phi _{\alpha_2} \dots \phi _{\alpha_n}
\rangle $. Then the equation for the nth moment can be written in the form
\begin{equation}
\sum_{\vbeta} ({\bf a}_{\valpha ,\vbeta} - {\bf G}_{\valpha ,\vbeta}) 
{\fy}_\vbeta ~=~
N_\valpha -{\partial{\fy}_\valpha\over\partial t}
\label{eq:matrix}
\end{equation}
where
\begin{equation}
{\bf a}_{\valpha,\vbeta} ~=~ -\sum_{i=1}^n A_{\alpha_i ,\beta_i} 
\prod_{{j \atop j\neq i}}^n \delta_{\alpha_j , \beta_j} ,
\label{eq:lambda}
\end{equation}
\begin{equation}
G_{\valpha,\vbeta} ~=~ \sum_{i=1}^n\sum_{\gamma=1}^N \Gamma_{\alpha_i
\gamma \gamma \beta_i} \prod_{{j \atop j\neq i}}^n 
\delta_{\alpha_j , \beta_j}+
\sum_{i,j \atop i<j}^n 
[\Gamma_{\alpha_i \beta_i \alpha_j \beta_j } +
\Gamma _{\alpha_i \beta_j \alpha_j \beta_i }] 
\prod_{k \atop k\neq i,j}^n \delta_{\alpha_k , \beta_k},
\label{eq:G}
\end{equation}
and
\begin{equation}
N_\valpha ~=~ D \sum_{i,j \atop i < j}^n \delta_{\alpha_i,\alpha_j}
\phi_{\alpha_1,\dots\alpha_{i-1},\alpha_{i+1},\dots\alpha_{j-1},\alpha_{j+1},
\dots\alpha_N}
\label{eq:N}
\end{equation}
This involves moments of order $n-2$.
\par
Eqn. (\ref{eq:matrix}) provides a convenient framework to analyze
the general properties of moments. It is of the form of a matrix
equation of order $N^n$. Using this, it will be shown 
that these moments diverge at sufficiently high order if the symmetry
condition, mentioned above, for $\bf M$ holds and therefore
the PDF of the $\phi$'s has a power law tail.

When there is no additive noise, that is $\bf N~=~0$, the Lyapunov
exponent $L(n)$ is given by the largest eigenvalue of the matrix
${\bf G}-{\bf a}$. 

\subsection{Divergence of moments}
\label{sec:div}
With finite additive noise in steady state, Eqn. (\ref{eq:matrix}) 
can be rewritten
\begin{equation}
\fy_\valpha ~=~
-\sum_{\vbeta} {{( {\bf G}- {\bf a}) }^{-1}}_{\valpha ,\vbeta}
\bf N_\vbeta 
\label{eq:inverse}
\end{equation}  
The objective here is to show that 
there exists some finite $n$ beyond which the moments do not exist.
Generically this will occur when the highest eigenvalue of ${\bf G}-{\bf a}$
passes through zero. Positive eigenvalues show that the solution has become 
ill defined, and that the moments have ceased to exist.  
Note that if the nth moments exist, the
moments of order $n-2$ must also exist, and hence the right hand side
of eqn. (\ref{eq:inverse}) exists. Therefore
one accomplishes the above objective by 
showing that the inverse matrix on the right
hand side of eqn. (\ref{eq:inverse}) has eigenvalues that must pass from
negative to positive, as n is increased. 

From the end of section \ref{sec:moments} , the value of $n$ where this
occurs corresponds to the point where $L(n)$ passes through zero. This
condition is in agreement with the result obtained through the
heuristic argument of the preceding publication~\cite{one}.

Thus to show this divergence, one must find the dependence of
the eigenvalues of ${\bf G}-{\bf a}$ on $n$, the order of the
moments. One must show that at least one of the eigenvalues of 
this matrix become positive for sufficiently large n.
\par
Note that the eigenvalues of ${\bf G}-{\bf a}$ will span some finite
domain $E_{min}$ to $E_{max}$ both of which are difficult to ascertain.
However one can find a number between these two extremes by forming
the scalar product 
$\hat{\bf x} \cdot (\bf G - a) \cdot \hat{\bf x}$, 
where $\hat{\bf x}$ is any unit vector.  
If this scalar product becomes positive, then at least one eigenvalue
is greater than zero, and therefore the nth moment is divergent.
If one makes the choice $\hat{\bf x} \propto (1,1,\dots ,1)$
then the scalar product can be evaluated. At this point we restrict the
random matrices $\bf M$ to be either real symmetric or anti-symmetric.
\begin{eqnarray}
\sum_{\valpha,\vbeta} x_\alpha & ( &{\bf G_{\valpha ,\vbeta}}- a_{\valpha,\vbeta})x_{\beta}
 ~=~  \nonumber \\
&&{ n \over N} \sum_{\gamma\delta =1}^N A_{\gamma,\delta}
+ {n\over N} \sum_{\gamma =1}^N \langle (\sum_{\alpha =1}^N M_{\alpha,\gamma})^2\rangle
+2 {n(n-1)\over N^2} \langle (\sum_{\alpha \beta =1}^N M_{\alpha \beta})^2
\rangle
\label{eq:Msq}
\end{eqnarray}
Now we need to analyze the behavior of this expression as a function
of $n$. The first two summations have prefactors proportional to $n$,
where as the last term is proportional to $n(n-1)$ and must
be positive. Therefore this scalar product must become positive for
sufficiently large n.
\par
If we consider this scalar product as the $A_{ij}$'s are
varied one obtains a similar result.  
Suppose we start with $A_{ij}$'s that are sufficiently negative
that the nth moment exists.
As the $A_{ij}$'s become less negative, 
at some point the scalar product becomes positive. Therefore at
this point the moments $\fy _\valpha$ are ill defined.  
Therefore for a fixed order of moments n, the moments can
be made ill defined by varying the matrix $\bf A$.
\par
These arguments show that  for a system described by
eqn (\ref{eq:vector}),  there
must always be moments of the $\phi$'s that diverge if sufficiently high
moments are examined.  All of the moments of a given order are expected
to diverge at the same time, if the matrix $\Gamma$ is not transformable
to block
diagonal form. Block diagonal form should only occur when the system of
eqns. (\ref{eq:vector}), can be decoupled into completely separate subsystems.
Aside from this case, all elements of ${{({\bf G} - \bf a)}^{-1}}$ will 
become divergent at the same $n$. This means for example,  
that $\langle \phi _1^n \rangle$ should diverge at
the same $n$ as $\langle \phi _2^{n-3} \phi _3^3\rangle$. This argument
shows that the divergence in moments should take place somewhere
between the nth and (n+2)th moment. It is quite reasonable to assume that
the power law tail for the PDF will take on a
non-integral value. Note that the divergence is not dependent on the
matrix $\bf N$, which is a function of $D$ the
strength of the additive noise. Therefore the exponent of the power law 
should also be unaffected by the strength of the additive noise. 
However the steady state distribution becomes ill defined in the 
limit of zero additive noise as we saw in section \ref{sec:absence}. 
The power law tail should only depend on $\bf\Gamma$ and $\bf A$.

\section{Physical Example}
\label{sec:reaction}
Here the results of the previous section are used to understand and
make predictions about an interesting physical system, the convection
of a passive scalar field such as temperature in a random velocity
flow. We
will add the extra ingredient that the fluid is undergoing an
irreversible
chemical reaction $A\rightarrow B$. All molecules start off as type A
and the reaction is exothermic. In general one expects the rate
constant to
be temperature dependent. Over a limited temperature range this
dependence can be approximated as linear. It is easily seen{nicolis} 
that this
reaction adds a term $\alpha \phi$ to the right hand side of eqn.
(\ref{eq:passive}). The equation under consideration is
\begin{equation}
\partial_t \phi + {\bf v} \cdot \nabla \phi = d \nabla^2 \phi + 
\alpha\phi + \eta (x,t)
\label{eq:alpha}
\end{equation}
The last term is an additive random noise term, the
correlation function in fourier space 
$\langle|\hat\eta ({\bf k} ,t)|^2\rangle$
must go to zero as $k \rightarrow 0$
to ensure heat conservation in the absence of chemical
reactions.

It will
now be argued that this system exhibits a cross-over from an
exponential tail in the PDF, $P(T)$, to a power
law as $\alpha$ is increased.

First consider the case of no reaction, $\alpha ~=~ 0$ and
no additive noise.
In this case we will now see that $L(q)$ must asymptote to a constant
$\leq 0$. There are three ingredients in this argument.
First the Laplacian acts as a short distance
cutoff precluding variation of the temperature field on a
smaller scale.  To estimate this length scale we are assuming
that the velocity field has a small distance cutoff. In the case
of turbulence this would be Kolmogorov's smallest
eddy scale . Another scale of relevance is obtained by
comparing the first two terms on the right hand side
of eqn. (\ref{eq:alpha}).  This yields a scale $l_c =d/v_{rms}$.
As is usual with competition between convection and
diffusion, it is the latter that will dominate below and the
former that will dominate above the scale $l_c$\cite{fluid}.
Therefore at a sufficiently small length scale the variation
of $T$ should be smooth.  Second, in a closed
system or one with zero net heat flux into the system, the integrated
temperature is conserved.  Third if the the scalar field is initially
positive everywhere, it must
remain so under evolution of eqn. (\ref{eq:passive}).
These three observations, of {\it (a)} conserved
heat flux, {\it (b)} minimum length scale, and {\it (c)}
positivity of $\phi$, imply that in
steady state there will be a maximum that the temperature can
take. This would be when the entire non-zero, or non-average,
temperature field is concentrated in a spike of size the minimum
length scale.
Because there is a maximum to $\phi$, it cannot grow exponentially
and therefore $L(q) \leq 0$.

But in fact from common experience one knows that without
additive noise temperature variations will
decrease in time. Physically it is apparent,
that after some time an isolated system should come
to equilibrium at a constant temperature. The  form
of the decrease
in r.m.s. temperature fluctuations depends on the form of
the random velocity field. The decrease in fluctuations will
depend, for long times, on the boundary conditions.
If there is no heat flux in or out of the system, the
integrated temperature is conserved. With $\bf v =0$ the
process is purely diffusional, and
one expects an exponential decay in fluctuations of the temperature
field.  The addition of a random velocity field should not change
this conclusion. It should have the effect of mixing the system
faster and increasing the rate of decay. If the boundary conditions
are no longer insulating so that heat flows out of the system, the rate
of decay should decrease further.  The exact dependence is
not important for the conclusions here,
but only that in a finite size system and for a variety
of boundary conditions, $L(q) < 0$ for $q > 0$.

Now consider the effect of adding a term $\alpha \phi$ to the
right hand side of eqn. (\ref{eq:passive}) to make the
fluid reactive as described above.  By letting $\phi \rightarrow
\phi \exp (\alpha t)$, one eliminates the term $\alpha \phi$
from the equation. Therefore the term $\alpha \phi$ modifies
the the Lyapunov exponents with no reaction $L_0(q)$ to 
$L(q) = L_0(q) + \alpha q$.  
Because $L_0(q)$ must be convex, sufficiently large $\alpha$ 
must cause $L_0(q)+\alpha q$ to cross zero at finite $q$.

If we now consider the steady state behavior of $P(\phi )$ with
additive noise we recall  that because $L(q) \leq 0$ for $\alpha = 0$,
$P(\phi )$ for one component models~\cite{one}
must be an exponential or stretched exponential.
For sufficiently large $\alpha$, one should observe a cross over
to power law behavior.  The precise cross over behavior
cannot be predicted from this analysis, only that such a
cross over must take place. The next section confirms this cross over
numerically.

\section{Numerical Simulations}
\label{sec:numerics}
We will illustrate and confirm some of the results presented here by
means of computer simulation. Consider eqn. (\ref{eq:passive}) in one
dimension. In one dimension the incompressibility constraint cannot be
enforced and still obtain an interesting problem, so this constraint is
dropped. The equation is discretized spatially giving 
\begin{equation}
{\dot\phi_i} + (\phi_{i+1}v_{i+1}-\phi_i v_i) ~=~ d(\phi_{i+1} -2\phi_i
+ \phi_{i-1}) \label{eq:discrete} \end{equation} 
This equation is solved
with periodic boundary conditions, and it is clear that the total $\phi$
$\Phi \equiv \sum \phi_i$ is conserved. Since we are interested
in the fluctuations it is most convenient to take $\Phi = 0$. 
The simplest ``order parameter''
for this system is the total standard deviation. 
\begin{equation}
\phi^2 \equiv \sum_{i=1}^{N} {\phi_i}^2 
\end{equation}
This is to be distinguished from the $\phi$'s being evolved 
by eqn. (\ref{eq:vector}). In this section we will
denote the order parameter simply by $\phi$.
The  different $v_i$'s are taken to
be independent and Gaussian distributed. This equation is updated by
fourth order Runge Kutta.  16 lattice sites are used, and the
distribution $P(\phi )$ was computed by averaging. Fig. \ref{fig:falpha}
plots $(\ln P(\phi))/t$ versus $(\ln \phi)/t$ for two different times
t. According to eqn. (\ref{eq:plnphi}) the resulting function should be
independent of time, up to an overall vertical shift. The two curves
have been shifted relative to each other and lie on top of each other
within the error bars confirming the validity of the scaling assumed
in this paper. Note that the data for the longer time PDF  does
not extend down as far as for the shorter time. There is a problem
measuring the tails of the PDF because the number of data
points becomes very small. This one dimensional problem may not
have an $L(q)$ that is less than zero for positive $q$ as the
positivity of $\phi_i$ is no longer guaranteed. 

We next turn to the case where additive noise is included. We shall
illustrate the transition in behavior from an exponential tail in the
PDF to a power law tail in the case of a reactive
randomly stirred fluid. Adding a term $\alpha \phi_i$ to the right hand
of eqn.  (\ref{eq:discrete}) represents a one dimensional discretized
version of the reacting system described in section
\ref{sec:reaction}.
Periodic boundary conditions are still employed here but we start
with the condition that the sum over all sites of $\phi_i$ is zero.
In this case for a finite size system and $\alpha = 0$, $\langle \phi_i
\rangle$ will decrease exponentially. This is seen from the fact
that when the velocity field is zero, the system is diffusive and
$\phi_i$ decreases exponentially. When velocity is added, this does
not alter this result. 
From the previous section it was predicted that a system with 
sufficiently large $\alpha$ should have power law tails.

Fig. (\ref{fig:add}) shows $\phi$ versus $P(\phi )$
for different values of $\alpha$. If $\alpha$ is too large all the
$L(q)$ are positive and the system is unstable. For $\alpha$ too
small there is no discernible power law behavior. However
this cannot be ruled out based on the data. There is a
narrow region where clear power law tails are manifest.

The method just used for obtaining the PDF has the draw back
of not being able to probe the tails of the distribution.
Equivalently, the high order moments and their $L(q)$'s are
difficult to obtain. It is the convergence of high 
order moments that presents a problem. Multiplicative
noise equations like eqn. (\ref{eq:vector}) have highly intermittent
behavior. This is a result of the fact that different moments
have different Lyapunov exponents. Therefore the typical behavior
of a high moment will be very different from its true average.
Flucutations that are exponentially unlikely in time dominate the
correct answer. Therefore straightforward 
averaging is practically limited to  
very low $q$, before the number of number of runs that are needed
to obtain convergence becomes far too large.

A way around this problem is to use a ``replication algorithm''.
Similar methods have been used in quantum simulations for the
past fifteen years~\cite{ceperley}.
In order for this method to work, one needs the order parameter $\phi$ to be
positive definite, which it is in most situations, as in the above
example.
A large number of copies of the system described by eqn. (\ref{eq:vector}) 
are made. These are
all updated to the next time step, with independently chosen random numbers,
the matrices $\bf M$.
A weight $w$ is computed by taking the ratio $w=\phi^n(t+\Delta t)/\phi^n(t)$.
Again $\phi$ can be taken to be the total $\phi$ field, or the standard
deviation of it, as above.  Using this total rather than $\phi_i$ at a
single point $i$
leads to better convergence and is therefore what is done in
practice. Once the weights $w$ are determined, the integer part of
w is taken and that number of copies is made. The remainder
of w is taken into consideration, by generating a random number
between 0 and 1 and making an additional copy if the random
number is smaller than the remainder. Additionally, to stop the
exponential explosion in the number of copies, the
weights are all normalized so that the number of copies is
kept almost constant.

The above method takes into account the idea of {\it importance sampling}.
When one is interested in the average value of the
moment $\phi^{n+1}(t)$, this is not computed directly, but the average
of $\phi$ is computed. It is computed for a biased ensemble of configurations,
each one having a probability proportional to $\phi^n$ of appearing.
Therefore one is computing $\langle \phi^n \phi\rangle /\langle \phi^n \rangle$.
This method is most powerful with no additive noise. In this case
the exponential divergence of this ratio is computed for long times, which
gives $L(n+1)-L(n)$.

The method is illustrated first for the two dimensional version of eqn. (\ref{eq:passive})
with no additive noise.
To make sure that $\nabla\cdot{\bf v} = 0$, $\bf v$ is derived from a vector
potential $\bf v = \nabla \times A$. 
The velocity field $\bf v$ is taken to be smooth but randomly varying in time.
$A$ is taken to be of the form 
\begin{equation}
{\bf A}(x,y,t) ~=~ M(t) \sin ( {2\pi (x-x_0(t))\over L}) \sin ({2\pi (y-y_0(t))\over L})
\hat z 
\end{equation}
New values of $x_0$ and $y_0$ are randomly chosen every $\Delta t$ steps.
$M(t)$ is a Gaussian random variable. This describes a circular motion
with periodic boundary conditions, moving with a random angular velocity.
There is no problem obtaining satisfactory convergence even for
large moments. The value of $\ln \phi$ as a function of time is plotted
for three different values of $q$ in fig. \ref{fig:phi(t)} for an $8 \times
8$ system..
Approximately 720 copies are run in this simulation. 
The slopes of these lines give
the difference in Lyapunov exponents $\Delta L \equiv L(q+1) -L(q)$.
This difference is plotted versus q in fig. \ref{fig:lqpassive} in a system 
of size $8\times 8$. The same quantities are plotted in fig. 
\ref{fig:lq2passive} in a system of size $16\times 16$. All other
parameters are kept the same and 180 copies were updated. 
Note that the variation of $L(q)$ is much smaller for this larger system.
Also note that $L(q)$ is always negative for positive $q$ and appears
convex. By adding a chemical reaction of the type described in section
\ref{sec:reaction} for sufficiently large $\alpha$, $L(q)$ should cross
zero leading to power law tails.

The final example considered here is the
population growth model, eqn. (\ref{eq:population}), 
with $\alpha = 0$. This has been
studied extensively because of its relevance to quantum
mechanical many body systems. The Lyapunov exponent $L(q)$
corresponds to the ground state energy of $q$ particles interacting
with each other via a two body potential as was first shown by 
Sugiyama and Koonin~\cite{koonin}.
Here we are considering a lattice model so that ${\bf r}$ takes discrete values
and $\nabla^2$ is the discrete Laplacian. Without the last
term, the above equation is just the diffusion equation, or the
imaginary time Schroedinger equation for a free particle. The
inter-particle interaction is a result of the multiplicative 
term as we shall see shortly. Assume that $f$ is Gaussian noise 
with zero mean and the correlation function 
$\langle f({\bf r},t)f({\bf r'},t')\rangle 
~=~ v({\bf r}-{\bf r'})\delta (t-t')$.

\def\br{{\bf r}}
For discrete $\bf r$ consider the nth moment averaged at
equal times
\begin{equation}
\Psi (\br_1,\br_2,\dots,\br_n,t) \equiv
\langle \phi(\br_1,t)\phi(\br_2,t)\dots\phi(\br_n,t)\rangle .
\end{equation}
In the notation of section \ref{sec:many}, $\Psi = \fy_\valpha$ and
$\vec\alpha$ corresponds to the coordinates $(\br_1,\br_2,\dots,\br_n)$
Translating eqns. (\ref{eq:matrix}-\ref{eq:G}) into these new
symbols it can be seen that
$\Psi$ obeys the many body Schroedinger equation with imaginary
time.
\begin{equation}
-\dot\Psi ~=~ H \Psi
\end{equation}
$H~=~ K+U$ is the Hamiltonian of the system which is the sum
of the kinetic and potential energy, $K$ and $U$ respectively
\begin{equation} 
K~=~ -\sum_{i=1}^n \nabla^2
\end{equation}
and
\begin{equation}  
U ~=~ -\sum_{i<j}^n u(\br_i-\br_j) - {n\over 2} u(0)
\end{equation}
Note that $u$ need not be positive as $f$ may be chosen to be
complex. This allows for the simulation of systems of particles
with both attractive and repulsive interactions. Also note that
the last term in the potential energy is constant and therefore
does not present any difficulties. The addition of a term 
$-U_{ext}(\br)\phi (\br )$
to eqn. (\ref{eq:population}) incorporates an external potential $U_{ext}$
in the Hamiltonian $H$.

Thus the wavefunction of a system of n bosons, with
certain initial conditions, is simply related to the nth moment 
of equation (\ref{eq:population}). To obtain the ground state
energy of n bosons, $E_n$, one need calculate the Lyapunov exponent
$L(n) = -E_n$. One is normally not interested in the constant
term $- {n\over 2} u(0)$ that appears in the expression for the
potential energy, so this is subtracted out.
Eqn. (\ref{eq:population}) has been studied numerically by path 
integral techniques~\cite{koonin}.
It is worth noting that the replication algorithm presented here
is also an efficient procedure for obtaining ground state energies.
Again, straightforward simulation of this equation will not
work because the higher order moments will not converge in a
reasonable time. The replication algorithm efficiently solves this
problem.

The method with the addition of the replication algorithm can
handle large number of particles and is quite computationally
efficient. Fig. \ref{fig:2dquantum} shows the difference in
energies $\mu_n \equiv E_n - E_{n+1}$ as a function of n
for a $10\times 10$ system with an on-site attractive 
potential $U=-0.5$. Approximately 900 copies are run in this simulation. 
In two dimensions there is a
transition between localized and delocalized particles as a 
function of density. The two dimensional transition can be
understood as a competition between kinetic and potential 
energy. The kinetic energy scales as $n/R^2$ where $R$ is
the correlation length.  The potential energy scales as
$u n(n-1)^2/(2\pi R^2)$. In the delocalized phase $R$ is the system
size $L$, and in the localized
phase $R$ should become as small as possible which is one
lattice spacing. This corresponds to all particles occupying
the same lattice site. The transition between localized and 
delocalized phases will then occur at approximately $n = 2\pi /u$.
For the figure shown the transition occurs within
a factor of two from the simple estimate given above. 
This transition is not a thermodynamic transition as it occurs at a
constant number of particles, almost independent of volume.

An interesting feature of the replication algorithm 
applied to this system is
that it exhibits metastability and hysteresis right above the transition. 
If the initial field $\phi (\br )$ is delocalized then it will stay so
for many iterations of the replication algorithm. It will eventually
jump to the localized phase and remain in that state. The interpretation
of this is similar to nucleation theory~\cite{ma}.

\section{Conclusions}

The PDF for eqn. (\ref{eq:vector}) has been analyzed 
for many components. It was found that PDF will exhibit
power law tails when a symmetry condition holds for
the random multiplicative matrix $\bf M$.

A passive scalar such as temperature in a random velocity
field is predicted to have a PDF with an exponential or
stretched exponential tail in agreement with earlier experimental
and theoretical work~\cite{kraichnan,libchaber,sreenivasan,ching,yakhot,chen,orzag,pumir}. 
If the fluid is reactive
this is predicted to become power law when the reaction
rate is sufficiently large. Numerical evidence was
presented supporting this prediction. The numerical
work suggests that this power law behavior is most easily 
seen in small systems where the velocity correlation length 
is comparable to system size.

It is interesting that the multi-component Gaussian case
should exhibit the same behavior as the 
a one component system albeit one that is no longer gaussian.
An argument can be given to understand this equivalence with
no additive noise.
A multidimensional system has a variety of relaxation times.
Consider the longest of these $T_{rel}$. For times much
longer than $T_{rel}$ a system looks 
statistically identical to how it did originally 
except for an overall shift in scale. To make this concrete,
suppose we shift the scale periodically after a time 
$\Delta T >> T_{rel}$ so that $|\vec\phi|^2 ~=~1$.
The shift in scale will not be completely deterministic
but will fluctuate because of the stochastic nature of 
the equation. For a long time $M \Delta T$,
the overall scale factor is obtained by multiplying 
the $M$ shifts in scale together. Each of these scale shifts
is random and in general is expected to be non-gaussian.
The nature of this non-gaussian distribution depends on the
evolution of the system within a relaxation time. 
Therefore the long time evolution of this system can still
be properly modeled by a one component system as was
investigated previously~\cite{one}

\acknowledgments
The author thanks Herbert Levine, Eliot Dresselhaus, Ken Oetzel,
and Matthew Fisher for useful discussions. This research
was supported by the NSF under grant DMR-9112767.

\begin{figure}
\caption{ The probability distribution of a field $\phi$ obeying eqn.
(15)
%(\ref{eq:passive}) 
in the absence of additive noise. The system
is in one dimension and is plotted with rescaled axis. The horizontal
axis is $\ln \phi /t$ and the vertical axis is $(\ln P)/t$. Two
different times $t_1 = 12.8$ and $t_2 = 38.4$ are shown.
The variance of the multiplicative noise is 2.0, the diffusion
coefficient d is $0.1$.
\label{fig:falpha}}
\end{figure}

\begin{figure}
\caption{The probability distribution of field $\phi$ obeying
eqn. (14)
%{(\ref{eq:passive})}
in one dimension, 
Six different values of $\alpha$ are shown.
The most steeply descending curve is for $\alpha ~=~ 0$, and
in order order of decreasing steepness, 
$\alpha ~=~0.40$, $\alpha ~=~0.47$, 
$\alpha ~=~0.49$, $\alpha ~=~0.50$, with the top curve at
$\alpha ~=~0.51$. The equation was discretized and the
number of lattice sites was chosen to be 8.
The variance of the both the multiplicative and additive 
noise is $0.6$.
\label{fig:add}}
\end{figure}

\begin{figure}
\caption{The logarithm of $\langle \phi\rangle$ as a function of time,
for a two dimensional passive scalr field.
The  average is a weighted average 
corresponding to the weights $\phi^1$ and $\phi^8$ and
$\phi^{16}$, calculated using
the replication algorithm described in the text. The slopes of these
lines correspond to $\Delta L \equiv L(q+1) -L(q)$, with $q=1,8$ and $16$.
The bottom curve corresponds to $q=1$, the middle curve to $q=8$,
and the top curve is for $q=16$
\label{fig:phi(t)}}
\end{figure}

\begin{figure}
\caption{The difference in Lyapunov exponent $\Delta L$ as a function
of $q$ for the two dimensional motion of a passive scalar field on
an $8\times 8$ lattice. Note
all of the values are monotonically decreasing confirming that $L(q)$
is convex.
\label{fig:lqpassive}}
\end{figure}

\begin{figure}
\caption{The difference in Lyapunov exponent $\Delta L$ as a function
of $q$ for the same system as the previous figure but on a $16\times 16$
lattice. Note the variation of $L(q)$ with $q$ is much smaller in this
larger system.
\label{fig:lq2passive}}
\end{figure}

\begin{figure}
\caption{The chemical potential at zero temperature,
 $\mu = E_n-E_{n+1}$ as a function of n, the number of bosons in a two
dimensional $10 \times 10$ lattice. The on-site attractive energy
is $-0.5$. The energies are calculated by the replication method
described in the text. Note the discontinuity at $n=19$.
\label{fig:2dquantum}}
\end{figure}

\newpage
\begin{figure}[ht]
\centering
\leavevmode
\epsfysize=20 cm \epsfbox{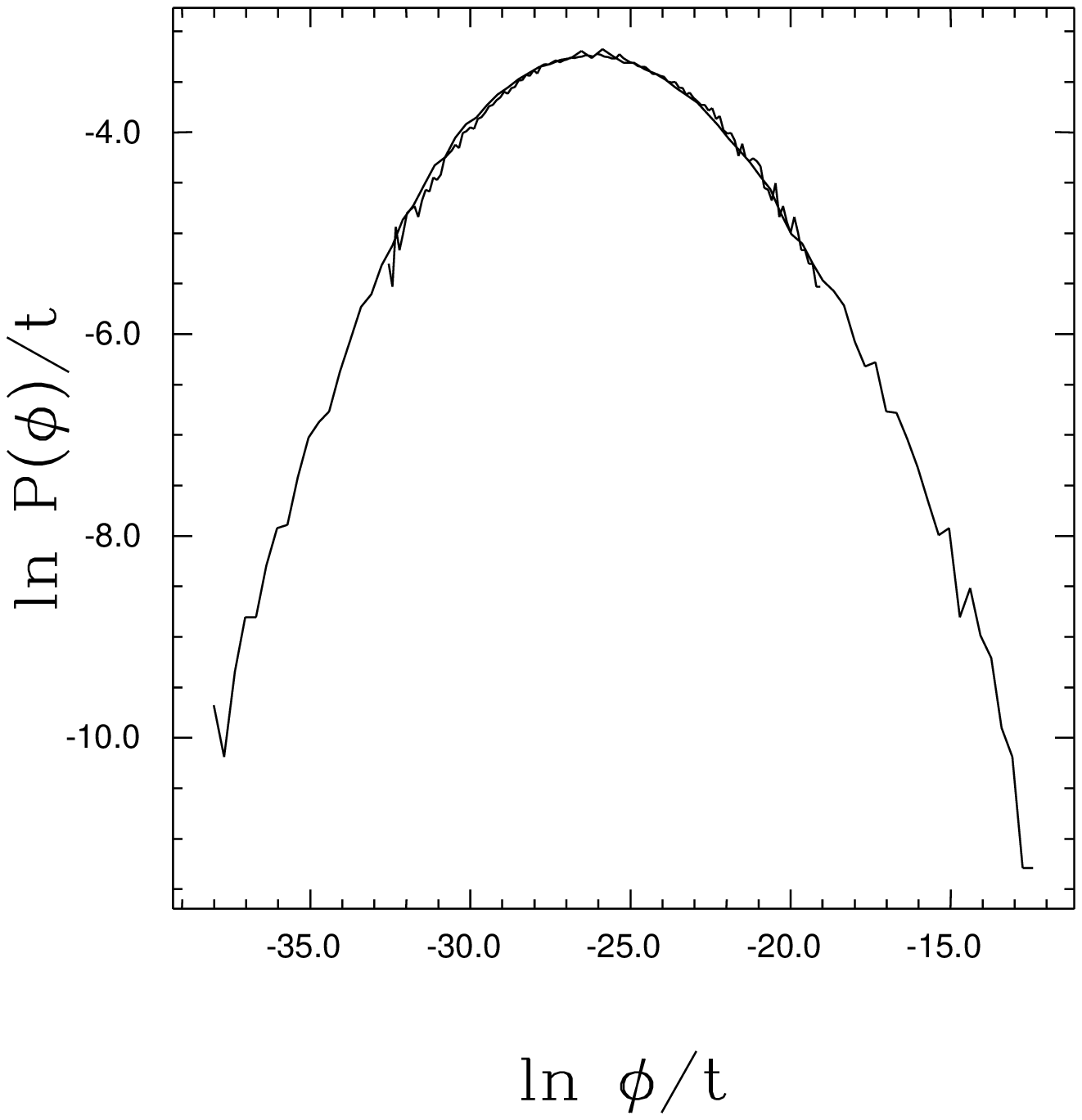}
\end{figure}
%\vskip 1in
\centerline{Fig. 1}
\newpage
\begin{figure}[ht]
\centering
\leavevmode
\epsfysize=20 cm \epsfbox{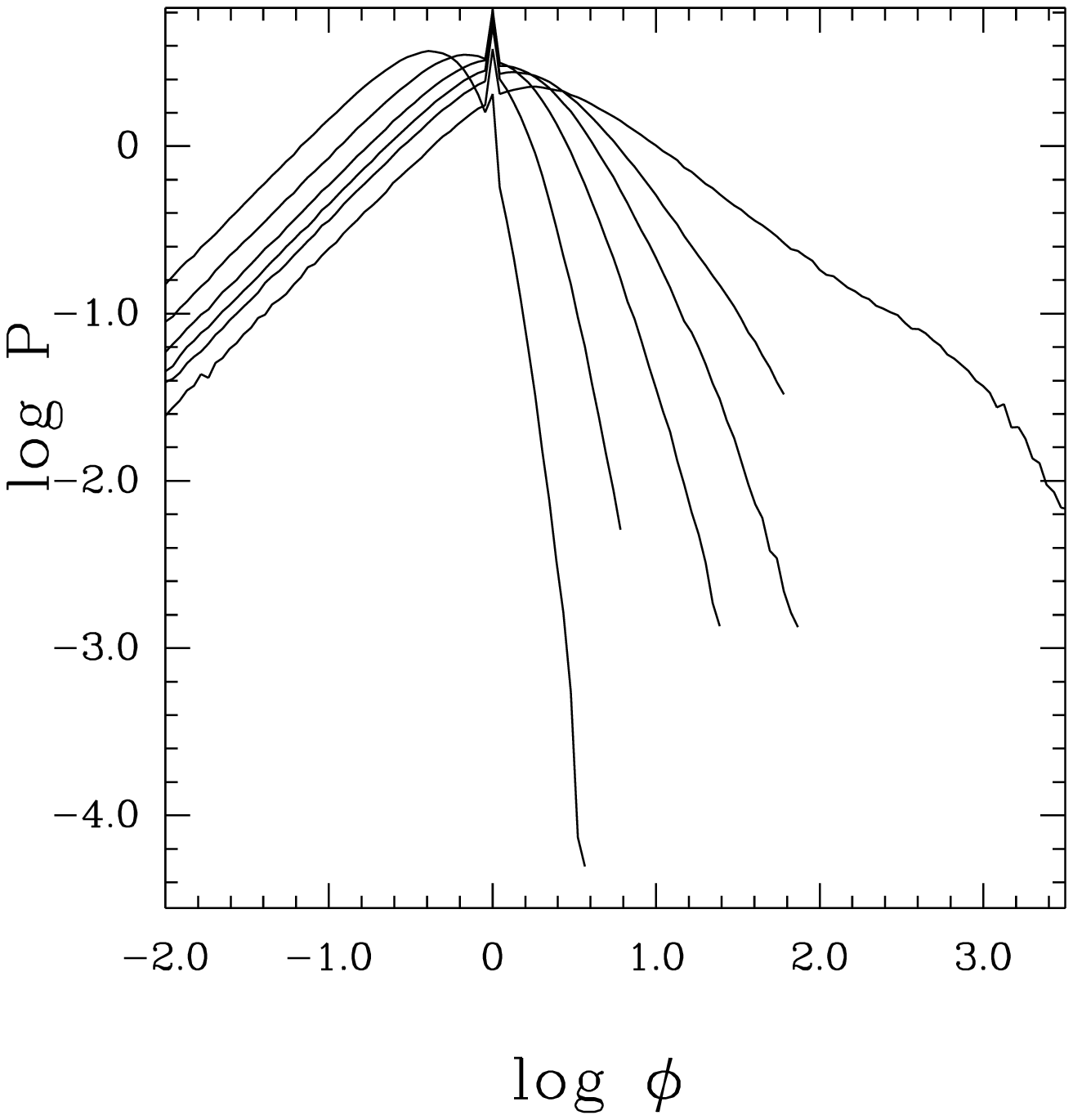}
\end{figure}
%\vskip 1in
\centerline{Fig. 2}
\newpage
\begin{figure}[htb]
\centering
\leavevmode
\epsfysize=20 cm \epsfbox{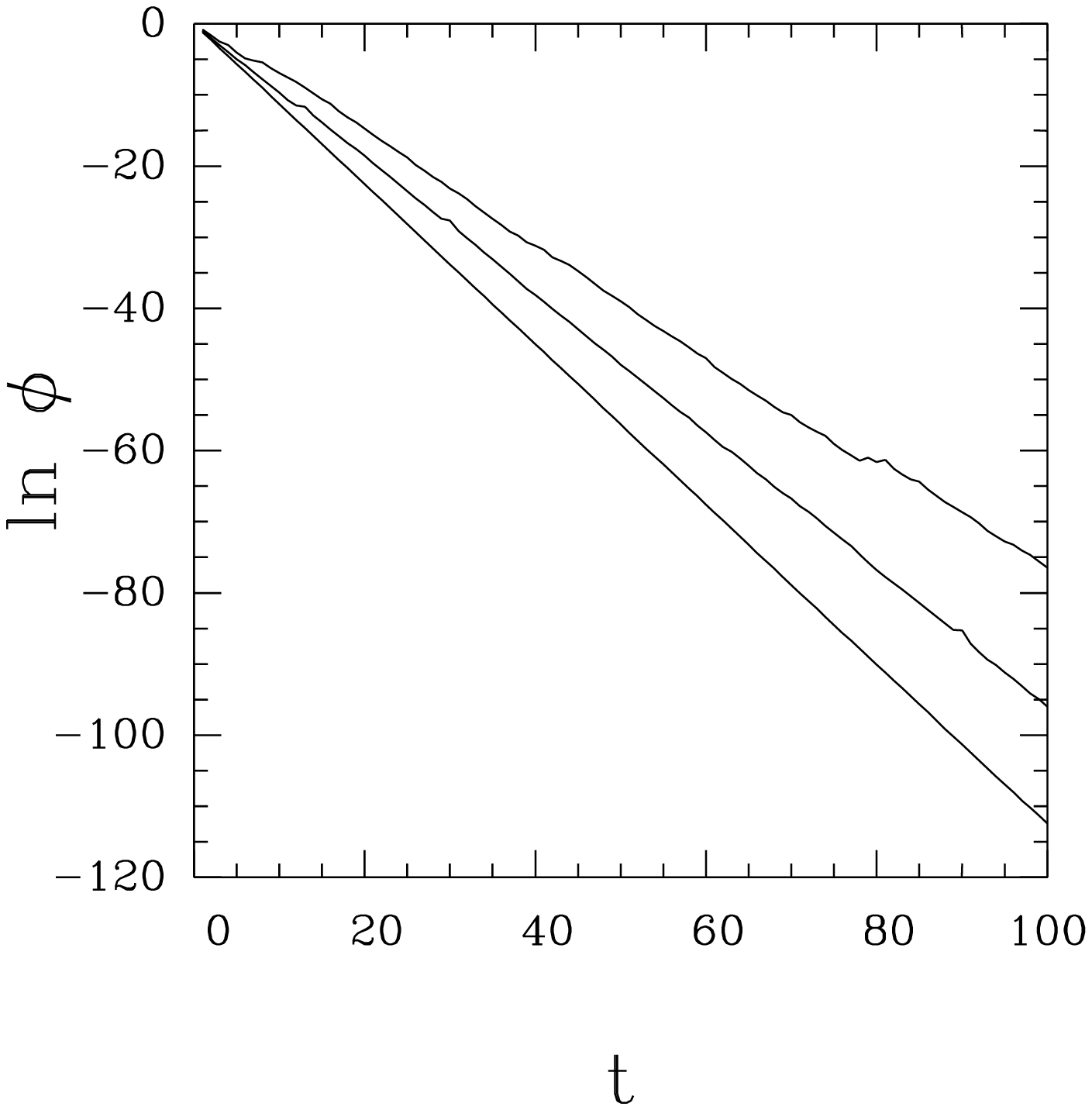}
\end{figure}
\centerline{Fig. 3}
\newpage
\begin{figure}[htb]
\centering
\leavevmode
\epsfysize=20 cm \epsfbox{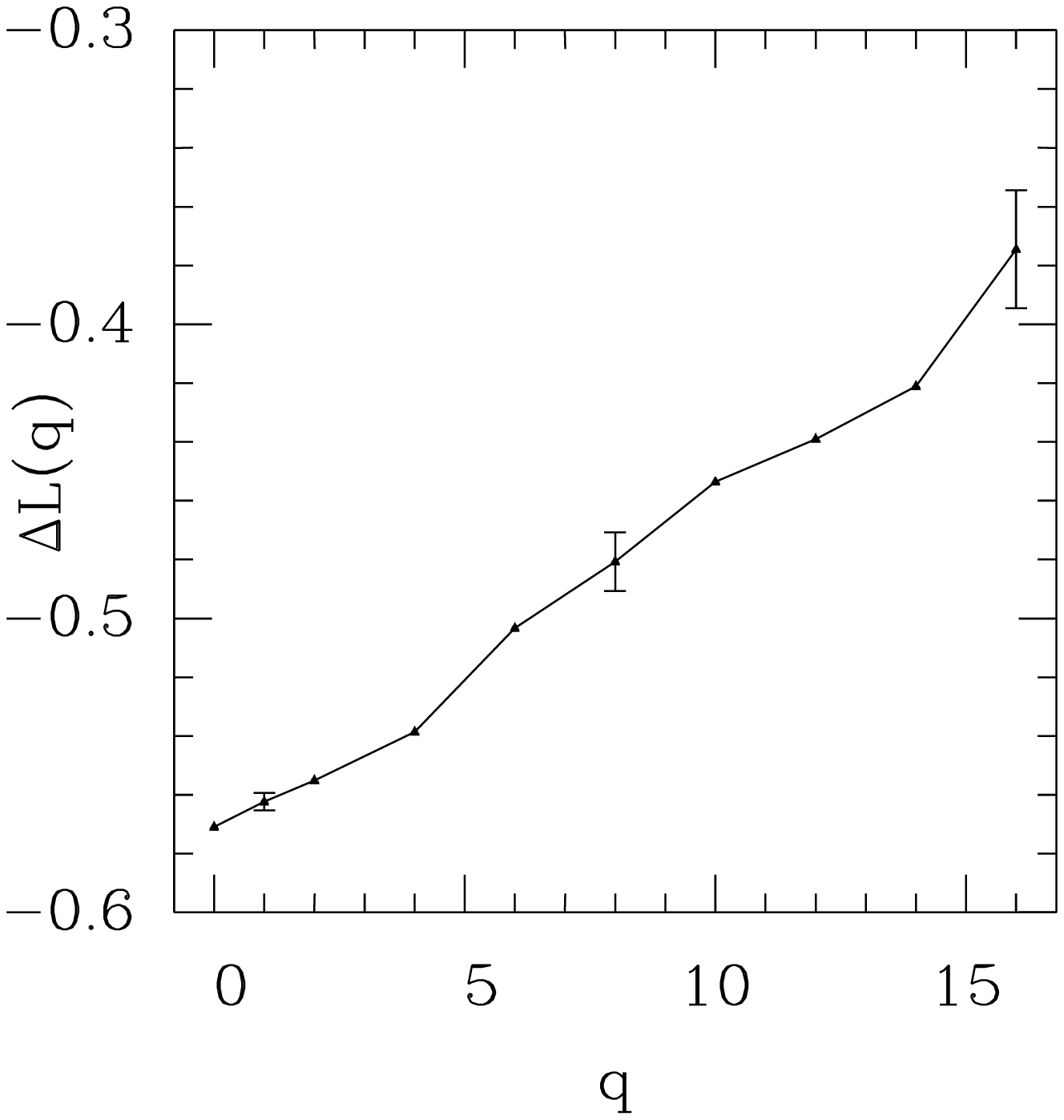}
\end{figure}
\centerline{Fig. 4}
\newpage
\begin{figure}[htb]
\centering
\leavevmode
\epsfysize=20 cm \epsfbox{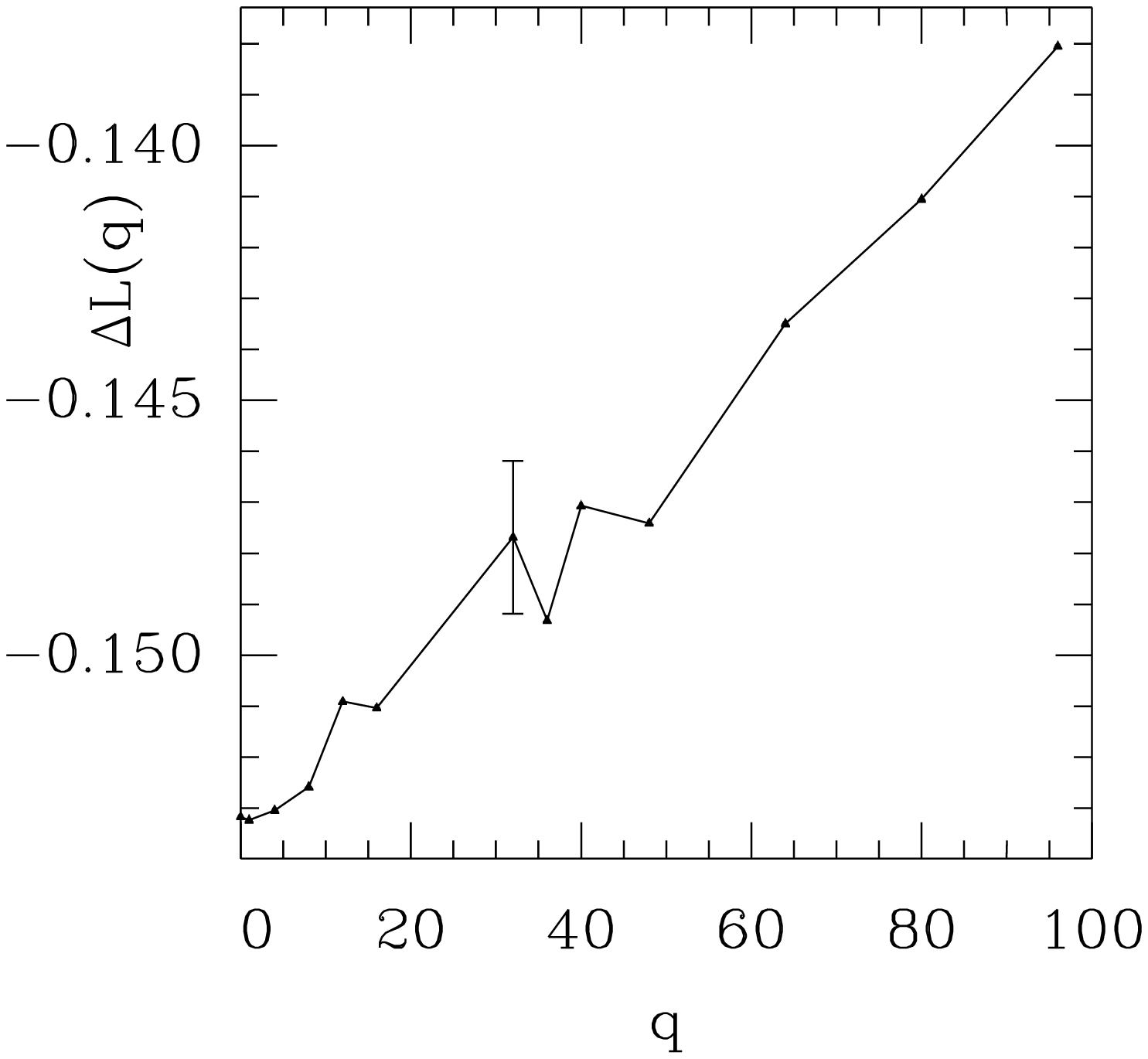}
\end{figure}
\centerline{Fig. 5}
\newpage
\begin{figure}[htb]
\centering
\leavevmode
\epsfysize=20 cm \epsfbox{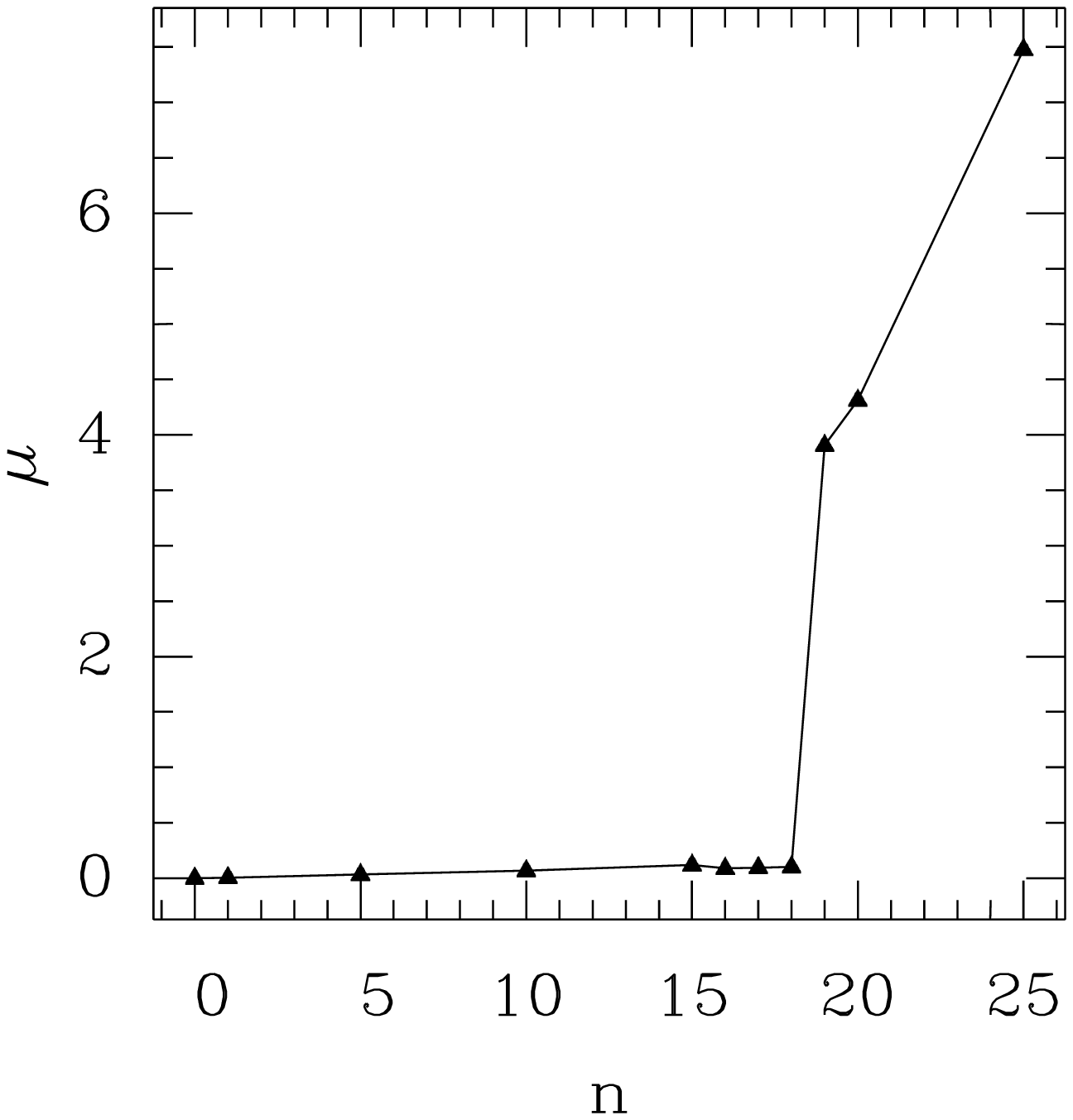}
\end{figure}
\centerline{Fig. 6}
\end{document}